\documentclass{osa-article}

\journal{oe}

\usepackage{float}

\begin{document}

\title{Tunable collective electromagnetic induced transparency-like effect due to coupling of dual-band bound states in the continuum}

\author{Jian Chen,\authormark{1,3,7} Rixing Huang,\authormark{2,3,7} Xueqian Zhao,\authormark{1,3} Qingxi Fan\authormark{3,4} Kan Chang,\authormark{2} Zhenrong Zhang,\authormark{1,5} and Guangyuan Li\authormark{3,4,6}}

\address{\authormark{1}Guangxi Key Laboratory of Multimedia Communications and Network Technology, School of Computer, Electronics and Information, Guangxi University, Nanning 530004, China\\

\authormark{2}School of Electrical Engineering, Guangxi University, Nanning 530004, China\\

\authormark{3}CAS Key Laboratory of Human-Machine Intelligence-Synergy Systems, Shenzhen Institutes of Advanced Technology, Chinese Academy of Sciences, Shenzhen 518055, China\\
\authormark{4}Shenzhen College of Advanced Technology, University of Chinese Academy of Sciences, Shenzhen 518055, China\\
\authormark{5}zzr76@gxu.edu.cn\\
\authormark{6}gy.li@siat.ac.cn\\
\authormark{7}These authors contributed equally.
}




\begin{abstract} The coupling between dual-band or multi-band quasi-bound states in the continuum (q-BICs) is of great interest for their rich physics and promising applications. Here, we report tunable collective electromagnetic induced transparency-like (EIT-like) phenomenon due to coupling between dual-band collective electric dipolar and magnetic quadrupolar q-BICs, which are supported by an all-dielectric metasurface composed of periodic tilted silicon quadrumers. We show that this collective EIT-like phenomenon with strong slow light effect can be realized by varying the nanodisk diameter or the tilt angle, and that the transparency window wavelength, the quality factor, and the group index can all be tuned by changing the nanodisk size. We further find that as the nanodisk size decreases, the slow light effect becomes stronger, and higher sensitivity can be obtained for the refractive index sensing. Interestingly, the sensitivity first increases exponentially and then reaches a plateau as the nanodisk size decreases, or equivalently as the group index increases. We therefore expect this work will advance the understanding of the collective EIT-like effect due to coupling between q-BICs, and the findings will have potential applications in slow-light enhanced biochemical sensing.
\end{abstract}

\section{Introduction}
Metasurfaces have emerged as a versatile platform in the field of optics and photonics and have found a diverse range of applications \cite{kildishev2013planar}. 
In many of these applications, achieving high quality factors ($Q$-factors) is crucial for the performance. As a simple way to achieve very large $Q$-factors, bound states in the continuum (BICs) have been the focus for enhancing light–matter interactions in metasurfaces and enabling diverse applications ranging from ultralow threshold lasing with directional emission, enhanced nonlinear effects, to ultrasensitive sensing \cite{NRM2016Marin_BICrev,SciBull2019Yuri_BICrev,AOM2021Kildishev_BICrev}.

Dual-band, and even multi-band quasi-BICs (q-BICs) and their applications, as well as their coupling effects have received increasing attention because of the rich underlying physics and extensive promising applications. In 2019, Cong and Singh demonstrated dual symmetry-protected BICs for normal incidence of $x$- and $y$-polarizations \cite{AOM2019dualBIC}. Later, dual-band BICs for the same polarization were theoretically reported \cite{NRL2021_BIC2,JO2020BIC2} and experimentally demonstrated \cite{AOM2022_dualBICplas,NP2022Li_dualBIC,JOSAB2024Li_dualBIC} using various structures. Taking advantage of dual-band q-BICs, 
enhanced emission \cite{AOM2020MagBIC2pl, NL2023Wang_dualBICemission}, enhanced harmonic generation \cite{OEA2022_dualBICNL,PRB2022Xiao_dualBICHHG}, 
and refractive index sensing \cite{ResPhys2022_dualBICsens,OE2023_dualBICsens,SciRep2023Hamidi_BICSens} have been demonstrated. Besides the applications, the weak or strong coupling between q-BICs is also attractive \cite{AP2019BICfanoCoupling,NP2020BICstrongcoupling,NP2021BICcoupling}. For example, the Kerker effects due to the coupling of two q-BICs were demonstrated for achieving Huygens' metasurfaces with tunable quality factors \cite{NL2018BIC_kerker}, phase-only modulation \cite{ACSP2020BIC_kerker}, and super-absorption \cite{ACSP2018BIC_kerkerAbs} or even perfect absorption \cite{COL2024BIC_kerkerAbs}. As another example, metasurface analogue of electromagnetically-induced transparency (EIT) was also numerically investigated \cite{OE2022BIC_EIT,OE2022BICcouplingEITkerker} or experimentally demonstrated \cite{LPR2021BIC_EIT}. Recently, some of the authors demonstrated collective EIT-like effect with extremely high quality factor and slow group index in all-dielectric metasurfaces based on two collective resonances of extremely narrow linewidths, one of which can even transfer to the symmetry-protected BIC at $\Gamma$ point \cite{NL2024Li_BIC_EIT}. Therefore, taking advantage of the Fano resonance between two ultra-narrow q-BICs, which provide a convenient approach to achieve ultra-narrow resonances, we would expect high-performance  EIT-like effect in terms of the quality factor and the group index.

All-dielectric metasurfaces composed of periodic quadrumers with C$_{4v}$ rotational symmetry support toroidal dipoles (TD) \cite{PRX2015_fourTD,RevPhys2020_fourTDrev} and thus have been of focus in the field of BICs. Different mechanisms of symmetry reduction were explored \cite{ACSP2018_fourTD,APL2023_fourdiskGroup} and further investigated using group theory \cite{JAP2019_fourdiskGroup,JPD2022_fourdiskGroup}. It was shown that both the toroidic and antitoriodic orders \cite{ANM2020_fourx3antiTD}, or both the ferromagnetic and antiferromagnetic orders \cite{PRA2020_fourAFM,JPD2020_fourAFM,SciRep2022_fourAFM,JOSAB2023_fourAFM,OM2024_fourAFMsens} can be excited simultaneously in such a simple system. Quite recently, some of the authors proposed a silicon quadrumer metasurface with displacement supporting dual-polarized dual-band q-BICs and demonstrated ultrahigh sensitivities in refractive index sensing and biosensing \cite{AOM2024Li_dualBICsens}. Results showed that under normal incidence of $y$ polarization, the dual-band q-BICs have a smaller spectral distance than those under $x$-polarization. However, by breaking the in-plane symmetry via displacement, these two q-BICs cannot spectrally overlap regardless of the displacement. Questions arise such as can these dual-band q-BICs be spectrally overlapped and result in EIT-like effect, and can this effect be tuned?  

In this work, we tackle these problems by proposing a silicon quadrumer metasurface with tilt rather than displacement to break the C$_{4v}$ symmetry. We will numerically show that dual-polarized dual-band q-BICs can also be supported, just like our previous work \cite{AOM2024Li_dualBICsens}. Interestingly, we will find that by varying the tilt angle or the nanodisk diameter, the dual-band q-BICs excited under the $y$-polarization can be tuned to be spectrally overlapped, resulting in the collective EIT-like effect. More strikingly, we will show that the transparency wavelength, the quality factor, and the associated slow light effect of the obtained collective EIT-like effect can be tuned by changing the nanodisk size. Finally, the role of the slow-light effect on the bulk sensitivity of the refractometric sensing application will also be discussed.

\section{Design and simulation setup}
\begin{figure}[!hbt]
\centering
\includegraphics[width=133mm]{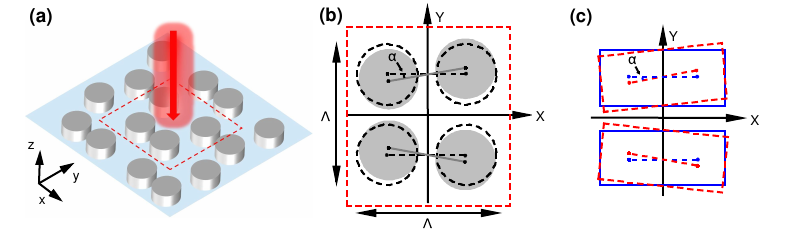}
\caption{(a)(b) Schematic diagram of the proposed silicon quadrumer array with periodicity $\Lambda$ in both $x$ and $y$ directions. The unit, as outlined by the dashed box, is composed of four nanodisks with diameter $d$ and height $h$. The top and bottom panels of nanodisk dimers are tilted with angle $\alpha$. (c) The proposed configuration is inspired by the metasurface of tilted nanodisks, which is widely adopted in BIC literature.}
\label{fig:schem}
\end{figure}

Figure~\ref{fig:schem}(a) illustrates the proposed silicon quadrumer array, the unit cell of which is composed of tilted nanodisk quadrumers. Unless otherwise specified, the nanodisks with diameter $d = 380$~nm, height $h = 200$~nm, and lattice period $\Lambda=1.1~\mu$m are embedded in a homogeneous dielectric environment of refractive index $n_0=1.45$, which can be guaranteed by covering the silicon nanodisks on a silica substrate $n_{\rm sub}=n_0$ with index-matching oil of $n_{\rm sup}=n_0$. The quadrumer of C$_{4v}$ symmetry within a unit cell reduces to C$_{s}$ symmetry by slightly tilting the top and bottom rows of nanodisk dimers at angle $\alpha$, as shown by Fig.~\ref{fig:schem}(b). Indeed, this configuration is inspired by the metasurface composed of periodic tilted-bar pairs, which is widely adopted in the field of BICs \cite{PRL2018Yuri_HighQBIC}, as illustrated in Fig.~\ref{fig:schem}(c). A novelty is that, each bar is now replaced by a row of two nanodisks. As we will show later, introducing a different configuration to break the in-plane symmetry can lead to q-BICs with distinct optical characteristics and physics.

We restricted ourselves to normal incidence of plane wave with electric field of unitary amplitude ($|E_0|=1$) and of polarization along the $x$- or $y$-direction. The zeroth-order transmittance spectra and the near-field distributions of the silicon metasurface are simulated with the home-developed rigorous coupled-wave analysis (RCWA) package, which was coded following \cite{JOSAA1995RCWA,JOSAA1997RCWA,PRB2006RCWA}. RCWA is a very powerful tool for simulating periodic photonic structures. In order to achieve the convergence region, we adopted a large enough order ($21\times 21$) \cite{OL2021Ku_HighQEQSLR}, and good agreement between the simulation results and the experimental data was reported previously \cite{AOM2024Li_dualBICsens}. To carefully estimate the resonance linewidths and quality factors, we adopted a small wavelength step of 0.01~nm. In all the simulations, the wavelength-dependent refractive indices of silicon were taken from \cite{PalikConstants}.

\section{Results and discussion}
\subsection{Collective EIT-like effect due to coupling of dual-band q-BICs}
\begin{figure}[!hbt]
\centering
\includegraphics[width=133mm]{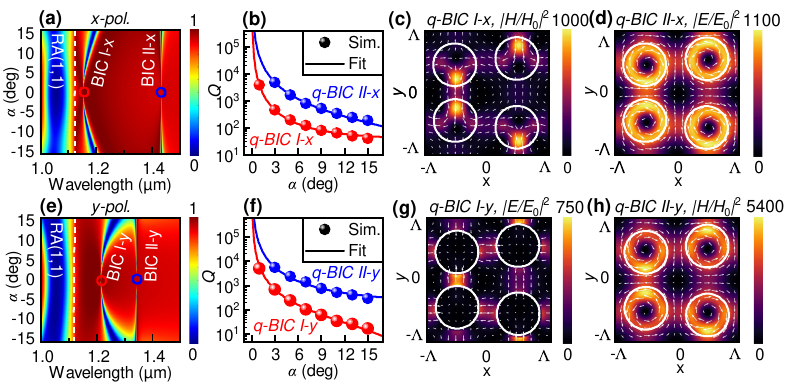}
\caption{(a) Simulated transmittance spectra of the silicon metasurface with different tilt angles under illumination of $x$-polarized plane wave. Circles indicate the occurrence of dual-band BICs at $\alpha=0^\circ$. Vertical white dashed lines indicate the RA wavelength of the $(\pm 1, \pm 1)$ order. (b) Extracted $Q$-factors of the dual-band q-BICs, which can be fitted with curves following inverse square relationships. (c)(d) Near-field magnetic or electric field ($|H/H_0|^2$ or $|E/E_0|^2$) distributions (color for intensity and arrows for directions) for these q-BICs corresponding to the transmittance dips for $\alpha=6^\circ$ in (a). (e)--(h) Similar to (a)--(d) but for $y$-polarization.}
\label{fig:xyPol}
\end{figure}
We first consider $x$-polarization. Fig.~\ref{fig:xyPol}(a) shows that there exist three pronounced resonances in the simulated transmittance spectra within the wavelength range of interest. The first resonance locates on the left-side and the second and the third resonances locate on the right-side of the RA wavelength of the $(1,1)$ order, which is given by
\begin{equation}
\label{eq:RA}
\lambda _{(1,1) {\rm RA}}= n_0 \varLambda/\sqrt{2}\,.
\end{equation}
As the tilt angle decreases, the second and the third resonances gradually narrow down and finally disappear when $\alpha=0^\circ$, at which the configuration is of C$_{4v}$ symmetry. We extracted the linewidths of these two resonances and obtained the $Q$-factors, which are estimated using the ratios of the resonance wavelengths to the linewidths, as function of the tilt angle. Fig.~\ref{fig:xyPol}(b) shows that the $Q$-factors of these two resonances can be fitted with the inverse-quadratic relationship, following
\begin{equation}
Q \propto 1/\sin^2\alpha\,.
\label{eq:QvsAng}
\end{equation}
The spectral and $Q$-factor characteristics suggest that the second and the third resonances are symmetry-protected q-BICs \cite{PRL2018Yuri_HighQBIC}, which transfers to BICs when $\alpha$ approaches 0. Note that in practice the $Q$-factors obtained in experiments will be limited by the array number \cite{SciRep2023Hamidi_highQArraySize,NC2021Boyd_HighQ} and the excitation source used \cite{NC2021Boyd_HighQ}, as well as by the fabrication imperfections \cite{AOM2024Li_dualBICsens}.

In order to understand the physics of the dual-band BICs, we examine the corresponding near-field distributions for $\alpha=6^\circ$. Figs.~\ref{fig:xyPol}(c)(d) show the near-field magnetic and electric field maps in the $x-y$ plane at the half height of the nanodisk for q-BICs I-$x$ and II-$x$, respectively. It is clear that the q-BIC I-$x$ features a collective in-plane magnetic dipole, and the q-BIC II-$x$ is the antiferromagnetic order.  

Under the $y$-polarised normal incidence, Fig.~\ref{fig:xyPol}(e)-(h) shows similar results, except that the electric field distribution of the q-BIC I-$y$ is similar to the magnetic field distribution of the q-BIC I-$x$, and the magnetic field of the q-BIC II-$y$ resembles the electric field of the q-BIC II-$x$. The near-field features suggest that the BIC I-$y$ should originate from the cancellation of anti-phase collective in-plane electric quadrupoles (EQs) that are supported by the two hybridized lattices, and that the BIC II-$y$ of the antitoroidic order should emerge from the destructive interference of anti-phase collective out-of-plane EQ.

Therefore, we have numerically shown that dual-polarized dual-band q-BICs are supported by the proposed silicon quadrumer metasurface with tilt. In general, most results are similar to our previous work \cite{AOM2024Li_dualBICsens}, except that the in-plane symmetry of the quadrumer is now broken by tilting the top and bottom rows rather than by displacing the diagonal nanodisks. As a consequence, an interesting and new finding is that, in Fig.~\ref{fig:xyPol}(e) the spectral distance between the q-BIC I-$y$ and the q-BIC II-$y$ decreases significantly as $\alpha$ increase. This is distinct from the results for the $x$-polarization in Fig.~\ref{fig:xyPol}(a), as well as those for both the $x$- and $y$-polarizations in our previous work \cite{AOM2024Li_dualBICsens}. 

\begin{figure}[!hbt]
\centering
\includegraphics[width=133mm]{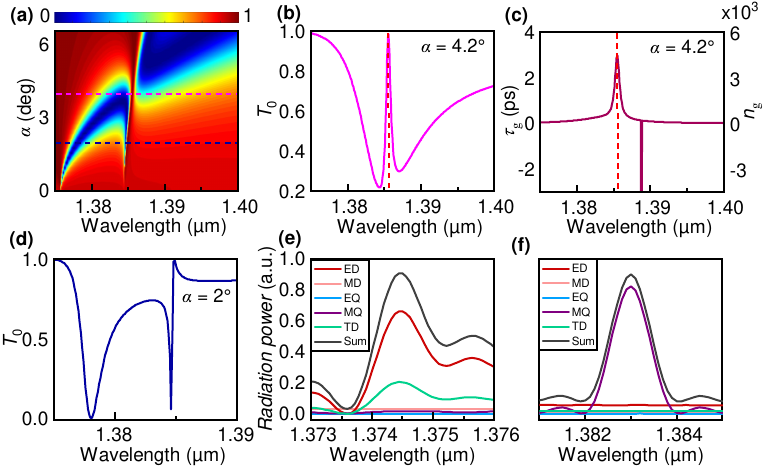}
\caption{(a) Simulated transmittance spectra as function of the tilt angle for given nanodisk diameter of $d=460$~nm. Vertical pink and black dashed lines indicate  $\alpha=4.2^\circ$ and $\alpha=2^\circ$, respectively. (b) Transmittance spectra, and (c) group delay/index spectra for $\alpha=4.2^\circ$. (d) Transmittance spectra for $\alpha=2^\circ$, and (e)(f) the multipolar contributions for radiation powers of q-BICs I-$y$ and II-$y$. ``ED'', ``MD'',  ``EQ'', ``MQ`` and ``TD'' stand for the electric and magnetic dipole, electric and magnetic quadrupole, and toroidal dipole moments. }
\label{fig:RTvsTilt}
\end{figure}

Although in Fig.~\ref{fig:xyPol}(e) the q-BICs I-$y$ and II-$y$ cannot merge because the left-top and the left-bottom nanodisks contact when $\alpha=15.8^\circ$, the spectral overlap can be realized by increasing the nanodisk diameter. For a larger nanodisk diameter of $d=460$~nm, Fig.~\ref{fig:RTvsTilt}(a) shows that these two q-BICs can spectrally overlap at $\lambda=1.3856~\mu$m when the tilt angle is $\alpha=4.2^\circ$, as indicated by the pink dashed line. In this scenario, a near-unitary transparency transmittance can be obtained, as shown by Fig.~\ref{fig:RTvsTilt}(b). The evolution of the Fano-shaped transmittance spectra as a function of the tilt angle indicates the occurrence of the EIT-like effect at $\alpha=4.2^\circ$. 

The EIT-like effect is accompanied by the slow light effect. The group delay is calculated by $\tau_{\rm g}=- {\rm d} \psi / {\rm d} \omega$ with $\psi$ the transmittance phase shift and $\omega$ the angular frequency, and correspondingly, the group index is calculated by $n_{\rm g}= c/v_{\rm g} = c \tau_{\rm g}/h$, where $c$ is the speed of light in vacuum and $v_{\rm g}$ is the group velocity of light. Fig.~\ref{fig:RTvsTilt}(c) shows that the calculated group delay and group index spectra. At the transparency wavelength, as indicated by the vertical dashed line, the group delay reaches as large as $\tau_{\rm g}=2.972$~ps, and the corresponding group index reaches up to $n_{\rm g}=4.464\times 10^3$. This indicates that the speed of light is slowed down by 4464 times in the silicon nanodisk quadrumers compared with in the vacuum. Compared with the literature, this group index is almost one third of the collective EIT-like slow-light effect based on the electric dipole and EQ collective resonances \cite{NL2024Li_BIC_EIT}, but is 45 times of the conventional EIT-like slow-light effect in a cavity-integrated guide-mode resonance grating \cite{JOSAB2021Li_tunableEIT}.

In order to understand the origins of this collective EIT-like effect, we perform multipolar decomposition for the two q-BICs at $\alpha=2^\circ$, which corresponds to the two transmittance dips in Fig.~\ref{fig:RTvsTilt}(d). The expressions for the Cartesian multipole moments' contributions can be found in \cite{ACSNano2018_decomp,PRB2019_decomp}. Figs.~\ref{fig:RTvsTilt}(e)(f) show that for the q-BIC I-$y$, the dominant contributing multipolar mode is the electric dipole (ED), whereas that for the q-BIC II-$y$, the dominant
mode is the magnetic quadrupole (MQ). This is because as the tilt angle increases, the gaps between the left two nanodiks and between the right two nanodisks both decrease, resulting in net dominant ED for the q-BIC I-$y$, consistent with the near field distribution in Fig.~\ref{fig:xyPol}(g). Similarly, for the q-BIC II-$y$, the net moment of the circulating magnetic fields as shown by Fig.~\ref{fig:xyPol}(h) is the MQ.

\begin{figure}[!hbt]
\centering
\includegraphics[width=133mm]{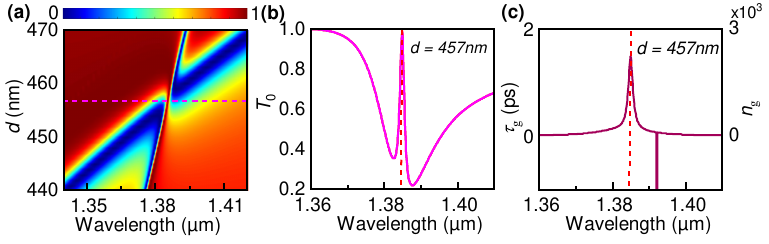}
\caption{Similar to Fig.~\ref{fig:RTvsTilt}(a)--(c) but for given $\alpha=6^\circ$. (b) and (c) are for $d=457$~nm, as indicated by the horizontal dashed line in (a).}
\label{fig:Tvsd}
\end{figure}

Alternatively, given the tilt angle, the q-BIC I-$y$ and the q-BIC II-$y$ can also be tuned to be spectrally overlapped for achieving the collective EIT-like effect by varying the nanodisk diameter. Fig.~\ref{fig:Tvsd}(a) shows that for $\alpha=6^\circ$, the two q-BICs spectrally overlap at the wavelength of 1.385~$\mu$m when $d=457$~nm. This overlap also results in near-unitary transparency transmittance, accompanied by slow light effect with $\tau_{\rm g}=1.496$~ps and $n_{\rm g}=2244$, as shown by Fig.~\ref{fig:Tvsd}(b)(c).  

\subsection{Turning the collective EIT-like effect via nanodisk size}

\begin{figure}[!hbt]
\centering
\includegraphics[width=133mm]{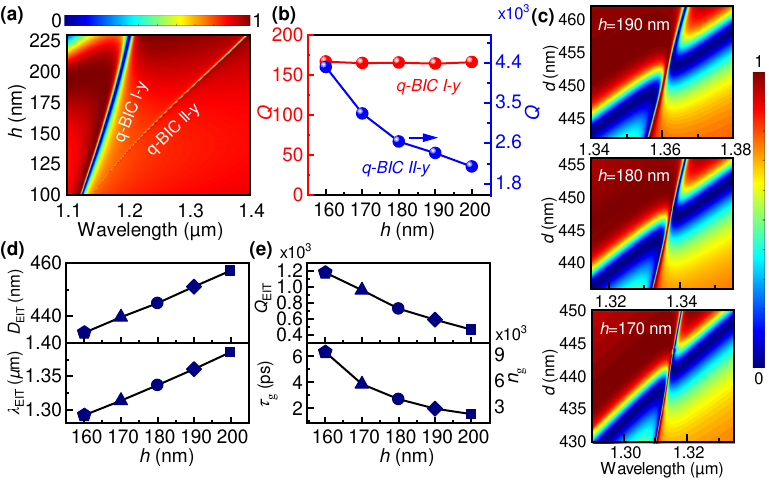}
\caption{(a) Simulated transmittance spectra for different nanodisk heights under $y$-polarization. (b) Extracted $Q$-factors for q-BICs I-$y$ and II-$y$ as functions of $h$. (c) Simulated transmittance spectra for different nanodisk heights and diameters. (d) The diameter and wavelength for achieving the collective EIT-like effect as functions of $h$, and (e) the associated $Q$-factors and the group delays/indices.  All the calculations were performed with $\alpha=6^\circ$.}
\label{fig:RTvsh}
\end{figure}

Figure~\ref{fig:RTvsh}(a) shows that as the nanodisk height $h$ decreases from 225~nm to 100~nm, both the q-BIC I-$y$ and II-$y$ are blue-shifted. Meanwhile, the linewidth of the q-BIC I-$y$ decreases slightly, whereas that of the q-BIC II-$y$ keeps decreasing. As a result, as $h$ decreases the $Q$-factors of the q-BIC I-$y$ remains almost constant, whereas that of the q-BIC I-$y$ increases dramatically, as shown by Fig.~\ref{fig:RTvsh}(b). This is because for the q-BIC I-$y$ the electric field is mainly confined between the nanodisks, and thus its $Q$-factor is almost independent from the nanodisk height; in contrast, for the q-BIC II-$y$ the electric field is mainly confined within the nanodisks, resulting in strong dependence on the height. In other words, varying the nanodisk height tune both the spectral distance and the $Q$-factor difference between the dual-band q-BICs excited under the $y$ polarization.

Therefore, given the tilt angle of $\alpha=6^\circ$, Fig.~\ref{fig:RTvsh}(c), together with Fig.~\ref{fig:Tvsd}(a), shows the tuning of the collective EIT-like effect via $h$. As $h$ increases, the nanodisk diameter for achieving the EIT-like effect should be scaled accordingly, resulting in linear increase of the wavelength for the transparency window, as shown by Fig.~\ref{fig:RTvsh}(d). Meanwhile, the $Q$-factor of the transparency window keeps decreasing, consistent with the behavior of the $Q$-factor for the q-BIC II-$y$. Correspondingly, the group delay/index also decreases with the nanodisk size, as shown by Fig.~\ref{fig:RTvsh}(e). This is because smaller $Q$-factor indicates shorter lifetime of photons, resulting in smaller group delay. Similar results can also be obtained for given the nanodisk diameter and varying the tilt angle, and are not presented here for the sake of simplicity. 

\begin{figure}[!hbt]
\centering
\includegraphics[width=133mm]{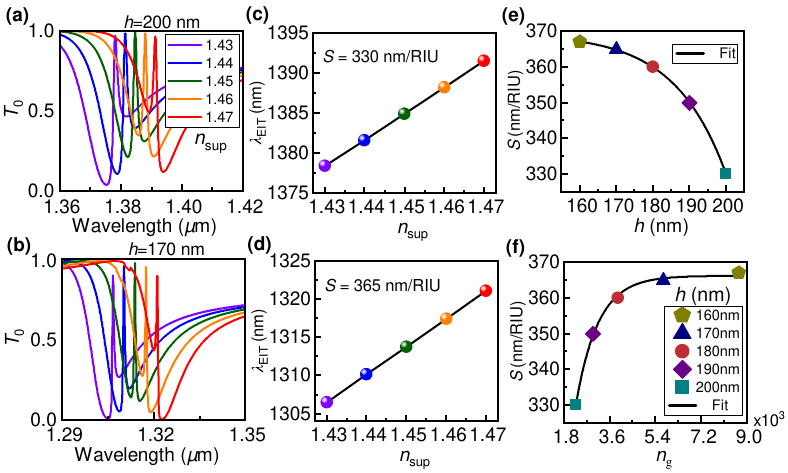}
\caption{(a)(b) Simulated transmittance spectra for height of $h=200$~nm or 170~nm with different superstrate refractive indices $n_{\rm sup}$. (c)(d) Corresponding transparency wavelengths as functions of $n_{\rm sup}$. The bulk sensitivity is the slope of linear fitting. (e)(f) Bulk sensitivity as a function of (e) nanodisk height or (f) group index.}
\label{fig:Sens}
\end{figure}

\subsection{Slow-light enhanced refratrometric sensing performance}
We now explore the potential sensing applications of the tunable collective EIT-like effect, and discuss the key role of the slow-light performance. We illustrate this with the tilt angle of $\alpha=6^\circ$. Fig.~\ref{fig:Sens}(a) shows that if $d=457$~nm and $h=200$~nm, the EIT-like transparency wavelength is redshifted linearly as the superstrate's refractive index $n_{\rm sup}$ increases from 1.43 to 1.47. By using linear fitting, the bulk sensitivity, which is defined as the ratio of the transparency wavelength shift to the refractive index change, $S\equiv \Delta \lambda_{\rm EIT}/ \Delta n_{\rm sup}$, is calculated to be $S=330$~nm/RIU, as shown by Fig.~\ref{fig:Sens}(c). Since the linewidth of the transparency window is only $\delta\lambda=2.97$~nm, the figure of merit (FOM), defined as the ratio of the bulk sensitivity over the linewidth, ${\rm FOM}=S/\delta \lambda$, is calculated to be FOM=111~RIU$^{-1}$.

If the nanodisk height is reduced to $h=170$~nm and correspondingly, the diameter is reduced to $d=440$~nm, similar behaviors can be observed in Fig.~\ref{fig:Sens}(b) but with relatively narrower linewidth of $\delta\lambda=1.37$~nm and larger spectral shift per refractive index unit. This corresponds to a relatively larger bulk sensitivity of $S=365$~nm/RIU and larger FOM of 266~RIU$^{-1}$, as shown by Fig.~\ref{fig:Sens}(d). 

We calculate the bulk sensitivities using the EIT-like windows for different nanodisk height and diameter pairs. Fig.~\ref{fig:Sens}(e) shows that as the nanodisk height decreases from 200~nm to 160~nm (the nanodisk diameter for achieving the EIT-like effect decreases accordingly), the bulk sensitivity increases exponentially from 330~nm/RIU to a plateau of 367~nm/RIU. This is because the bulk sensitivity depends on the fraction of the electric energy confined within the sensing volume, and can be expressed as
\cite{JOSAA2012Potyrailo_SensEq}
\begin{equation}
S=-2\frac{\omega_0}{n_{\rm a}} \frac{\int_{V_{\rm a}} \epsilon |E|^2 {\rm d} V}{\int_{V} \left(\epsilon(\omega_0) + \frac{\partial(\omega \epsilon)}{\partial \omega}|_{\omega_0} \right) |E|^2 {\rm d} V}\,.
\label{eq:Sens}
\end{equation}
The shorter the nanodisk height (correspondingly the smaller the nanodisk diameter), the larger the fraction of the electric energy confined within the sensing volume, resulting in the improved bulk sensitivity. We further plot the bulk sensitivity as a function of the group index in Fig.~\ref{fig:Sens}(f). Interestingly, as the group index increases, the bulk sensitivity also first increases exponentially and then gradually saturates around $S=367$~nm/RIU. This behavior clarifies the important role of the slow-light effect on the sensing performance of the proposed silicon metasurface.

\begin{figure}[!hbt]
\centering
\includegraphics[width=133mm]{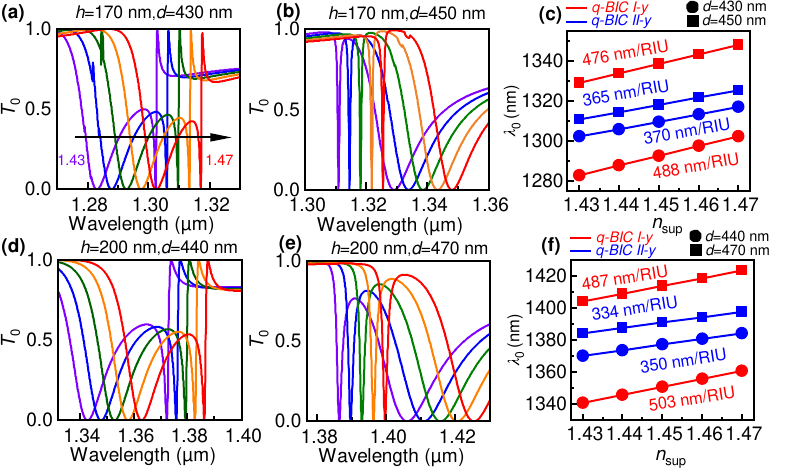}
\caption{(a)(b)(d)(e) Simulated transmittance spectra for (a) $h=170$~nm, $d=430$~nm, (b) $h=170$~nm, $d=450$~nm, (d) $h=200$~nm, $d=440$~nm, and (e) $h=200$~nm, $d=470$~nm for different superstrate refractive indices $n_{\rm sup}$ varying from 1.43 to 1.47 in step of 0.01. (c)(f) Corresponding resonant wavelengths for (c) $h=170$~nm and (f) $h=200$~nm versus $n_{\rm sup}$ with fitted slopes for bulk sensitivities.}
\label{fig:SensBIC}
\end{figure}

Although the bulk sensitivities of the EIT-like resonances are much larger than those of most q-BICs in all-dielectric metasurfaces \cite{OE2020Polyutov_highQDSensBIC,JAP2019Gil_highQDSensBIC,ACSN2020Zito_highQDSens, AOM2022Lin_highQDSensTD}, these results are smaller than that of the q-BIC I-$y$ in our previous work \cite{AOM2024Li_dualBICsens}. We further calculated the bulk sensitivities of the q-BICs I-$y$ and II-$y$ for the proposed configuration with various parameters. Figs.~\ref{fig:SensBIC}(a) and (b) show that, as the superstrate refractive index increases, the resonant wavelengths of the two q-BICs for $h=170$~nm and diameters before and after the spectral overlapping, respectively, are red-shifted. The resonant wavelengths as functions of $n_{\rm sup}$ in Fig.~\ref{fig:SensBIC}(c) reveals that the bulk sensitivities of the q-BICs I-$y$ and II-$y$ are approximate to $S=480$~nm/RIU and 365~nm/RIU, respectively, regardless of the diameter. Similarly, for $h=200$~nm, the bulk sensitivities of the q-BICs I-$y$ and II-$y$ are approximate to $S=490$~nm/RIU and 340~nm/RIU, respectively. These results are consistent with our previous work \cite{AOM2024Li_dualBICsens}. We also note that for different nanodisk heights, the bulk sensitivities of the EIT-like resonances are approximate to those of the q-BIC II-$y$ resonances. 

\begin{figure}[!hbt]
\centering
\includegraphics[width=133mm]{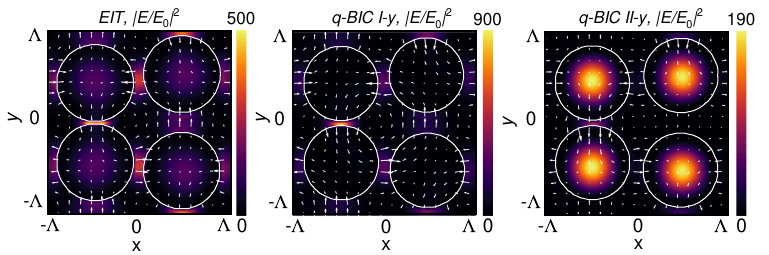}
\caption{Simulated near-field electric field distributions $|E/E_0|^2$ for (a) the EIT-like resonance with $d=457$~nm, and (b) the q-BICs I-$y$ and (c) II-$y$ with $d=440$~nm. The calculations were performed with nanodisk height of $h=200$~nm and tilt angle of $\alpha=6^\circ$.}
\label{fig:Fields}
\end{figure}

In order to understand the close bulk sensitivities for the EIT-like resonance and the q-BIC II-$y$, we turn to the near-field optical pictures. Fig.~\ref{fig:Fields} shows that the  electric field distribution of the EIT-like resonance can be regarded as the merging of those of the two q-BICs. As a result, a large portion of the electric field energy is confined within the nanodisks, just like the q-BIC II-$y$. According to Eq.~(\ref{eq:Sens}), the bulk sensitivities of the EIT-like resonance and of the q-BIC II-$y$ are close to each other. In other words, for the bulk sensitivity of the EIT-like resonance, the dominate contributions are from the q-BIC II-$y$, which has narrower linewidth. This is consistent with the $Q$-factor, as shown by Figs.~\ref{fig:RTvsh}(b)(e).

\begin{figure}[!hbt]
\centering
\includegraphics[width=133mm]{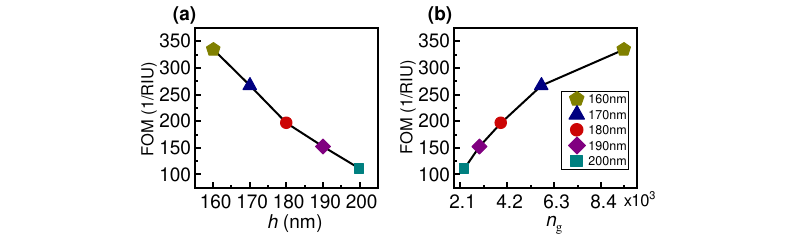}
\caption{Calculated sensing FOM of the EIT-like resonance versus (a) nanodisk height and (b) group index.}
\label{fig:FOM}
\end{figure}

Since the FOM can be expressed as 
\begin{equation}
{\rm FOM_{\rm EIT}} = \frac{S \cdot Q_{\rm EIT}}{\lambda_{\rm EIT}}\,,
\label{eq:FOM}
\end{equation}
we can plot the FOMs of the EIT-like resonances as functions of the nanodisk height and the group index by combining Figs.~\ref{fig:RTvsh}(e) and \ref{fig:Sens}(e)(f). Fig.~\ref{fig:FOM}(a) shows that, as the nanodisk height decreases from 200~nm to 160~nm (the diameter decreases accordingly for achieving the EIT-like effect), the FOM increases from 111~RIU$^{-1}$ to 333~RIU$^{-1}$. Correspondingly, the FOM increases with the group index, as shown by Fig.~\ref{fig:FOM}(b). We note that, these FOMs of the EIT-like resonances, which are also mainly determined by the q-BIC II-$y$, just like the bulk sensitivities and the $Q$-factors, are much larger than those of the q-BIC I-$y$. Such high FOMs may make the collective EIT-like resonance attractive in biochemical sensing applications.

\section{Concluding remarks}
In conclusion, we have proposed an all-dielectric metasurface composed of periodic tilted quadrumer array for tunable collective EIT-like effect. We have shown that the collective EIT-like effect, which is due to the coupling of the dual-band collective ED and MQ q-BICs, can be achieved by varying the nanodisk diameter or the tilt angle, and that the transparency wavelength, the quality factor, and the associated group delay or the group index of the EIT-like window can be tuned by changing the nanodisk size. Taking advantage of the tunable EIT-like window with stronger slow light effect for shorter nanodisks, we have found that the bulk sensitivity and the FOM of the refractometric sensing application can be improved dramatically. More interestingly, results have clarified that, as the group index increases, the bulk sensitivity increases exponentially and gradually reaches a plateau around 367~nm/RIU. We expect this work will enrich the physics underlying the coupling of dual-band q-BICs and find potential applications of slow-light enhanced sensing. As a final remark, we mention that dynamic tuning of the collective EIT-like resonance based on the coupling of two q-BICs can also be realized by incorporating active materials in the metasurfaces.

\begin{backmatter}
\bmsection{Funding}
National Natural Science Foundation of China (62275261); Natural Science Foundation of Guangdong Province (2024A1515011835); Guangdong Guangxi Joint Science Key Foundation (2021GXNSFDA076001); Guangxi Major Projects of Science and Technology (2020AA21077007).

\bmsection{Disclosures}
The authors declare no conflicts of interest.

\bmsection{Data availability} Data underlying the results presented in this paper are not publicly available at this time but may be obtained from the authors upon reasonable request.

\end{backmatter}

\bibliography{sample}






\end{document}